\documentclass[12pt,nofootinbib]{revtex4-1}
\usepackage{amsmath,amssymb}
\usepackage{slashed,verbatim,graphicx}
\usepackage{color}

\newcommand\vev[1]{\langle #1\rangle}
\newcommand\ket[1]{| #1\rangle}
\newcommand\bra[1]{\langle #1|}

\newcommand\BR{\mathbb{R}}

\newcommand{\beq}{\begin{equation}}
\newcommand{\beqs}{\begin{equation*}}
\newcommand{\eeq}{\end{equation}}
\newcommand{\eeqs}{\end{equation*}}
\newcommand{\CO}{  {\cal O}  }

\begin{document}
\setlength{\unitlength}{1mm}
\title{Conformal perturbation theory, dimensional regularization  and AdS/CFT}

\author{David Berenstein, Alexandra Miller}
\affiliation { Department of Physics, University of California at Santa Barbara, CA 93106}

\begin{abstract} 
We study relevant deformations of conformal field theory on a cylinder using conformal perturbation theory, and in particular the one point function of the deformation operator and the energy in a system after a quench. We do the one point function calculation in both AdS and the conformal field theory and we show that the results match. Our calculations are done with arbitrary spacetime dimension, as well as arbitrary scaling dimension of the relevant operator.  The only singularities that appear in the end calculation can be related to logarithmic singularities in dimensional regularization. We also study time dependent setups in the field theory and we show how the response of the system can be calculated in a Hamiltonian based approach. We use this procedure to explain certain short time universal results that have been found previously.

\end{abstract}

\maketitle

\section{Introduction }
\label{S:Introduction}

Consider a {c}onformal field theory in $d$ dimensions perturbed by a relevant (scalar) operator $\CO$ of dimension $\Delta<d$. 
We are interested in evaluating the correlators of $\CO$ in the presence of the perturbation.
The partition function is 
\begin{equation}
Z= \langle \exp( -\alpha\int d^d x  \CO(x) )\rangle= \left\langle \sum_{n=0}^\infty \frac 1{n!}\left(-\alpha\int d^d x  \CO(x) \right)^n\right\rangle\label{eq:Z}
\end{equation}
and the formal evaluation of correlators with the infinite sum in the equation above is what is known as conformal perturbation theory.
To begin with such a program, one can compute the one point function of $\CO(x)$ as follows
\begin{equation}
\vev{\CO(x)} {= \left\langle{-\alpha \int  d^d y  \CO(y) \CO(x)}+\dots\right\rangle} = -\alpha \int d^dy\frac{1}{|x-y|^{2\Delta}}{+\dots}\label{eq:onep}
\end{equation}
The right hand side is infinite regardless of $\Delta$. The divergence comes either from the small distance UV regime, or from the long distance IR regime. 
This is because we have to perform an integral of a scaling function. The problem seems ill defined until one resums the full perturbation expansion. This is a very
important conceptual point in the AdS/CFT correspondence \cite{Maldacena:1997re} where standard `experiments' insert time dependent or time independent sources for various fields on the boundary of $AdS$ \cite{Gubser:1998bc,Witten:1998qj} and these in turn 
can be associated {with}  sources for an operator such as $\CO(x)$. Some of these results have been argued to be 
universal in \cite{Buchel:2012gw,Buchel:2013lla,Buchel:2013gba,Das:2014jna}, independent of the AdS origin of such a calculation. We want to understand such type of results under a more controlled setting, where we can use the philosophy of conformal 
perturbation theory to get finite answers {\em ab initio} without a resummation.

A natural way to solve the problem above is to introduce a meaningful infrared regulator, so that the only divergences that survive arise from the UV of the theory and can then be handled via the usual procedure of 
renormalization.
Such a natural regulator is provided by the conformal field theory on the cylinder $S^{d-1}\times \BR$, which also happens to be the conformal boundary of global $AdS$ spacetime, rather than just the Poincar\'e patch.
The cylinder also is conformally equivalent to flat space and provides both the radial quantization and the operator state correspondence.
In this sense, we are not modifying the $AdS$ space in a meaningful way. However, a constant source for $\CO(x)$ in such a geometry is different than a constant source on the Poincar\'e patch.

In the rest of the paper we discuss the details of such a computation for two universal quantities. These are the one point function of $\CO(x)$, and the energy stored in a configuration where we quench from $\alpha\neq 0$ to $\alpha=0$. We also explain how to deal with general time dependent sources in the conformal field theory side for more general AdS motivated experiments.
 Because we work with arbitrary $d, \Delta$, our results can naturally be cast as a real space dimensional regularization formalism. 
 
 We find that the $AdS$ answer, which is generally finite for the one point function, matches this notion of dimensional regularization. The only singularities that arise are those that one associates {with} logarithmic divergences. We are also able to match this result to the CFT calculation exactly, where the calculation is more involved. We also argue how to calculate the energy of the configuration and that having solved for the one point function naturally produces the result for this other computation.

\section{One point functions on the sphere}

What we want to do is set up the equivalent calculation to \eqref{eq:Z} and \eqref{eq:onep}, but where we substitute the space $\BR\times S^{d-1}$ in the integral. {That} is, we want to compute
\begin{equation}
\vev{\CO(\tau, \theta)} \simeq \left\langle{-\alpha \int  d^{d-1} \Omega' d\tau'  \CO(\tau', \theta') \CO(\tau, \theta)}+\dots\right\rangle = -\alpha C_\Delta \label{eq:onepsp}
\end{equation}
for $\tau$ a time coordinate on $\BR$ and $\theta$ an angular position on the sphere. Because the operator $\CO$ is  not marginal, $\alpha$ has units and we need to choose a specific radius for the sphere. We will choose this radius to be one. Our job is to compute the number $C_\Delta$. Because the sphere times time as a space is both spherically invariant and time independent, properties that are also shared by the perturbation, we find that the result of equation \eqref{eq:onepsp} should be independent of both $\theta$ and $\tau$. As such, we can choose to perform the angular integral by setting the point $\theta$ at the north pole of the sphere, so that we only need to do an integral over the polar angle in $\theta'$. We want to do this calculation both in the AdS spacetime and in conformal field theory. We will first do the AdS calculation and then we will do the conformal field theory calculation.

\subsubsection{The AdS calculation}

As described in the introduction, we need to compute the answer in global $AdS$ spacetime. We first describe the global $AdS$ geometry as follows
\begin{equation}
ds^2=-(1+r^2) dt^2 +\frac {dr^2}{(1+r^2)} + r^2 d\Omega_{d-1}^2
\end{equation}
We need to find solutions for a perturbatively small scalar field $\phi$ with mass $m$ and time independent boundary conditions at infinity. Such a perturbation is a solution to the free equations of motion of the field $\phi$ in global $AdS$. Such boundary conditions allow separation of variables in time, angular coordinates and $r$. A solution which is time independent and independent of the angles can be found. We only need to solve the radial  equation of motion. Using $|g| \propto r^{2(d-1)}$ we find that we need to solve
\begin{equation}
\frac 1{r^{d-1}} \frac{\partial}{\partial r} \left(  r^{(d-1)} (1+r^2)\frac{\partial}{\partial r}  \right)-m^2 \phi(r)=0
\end{equation}
The nonsingular solution at the origin is provided by
\begin{equation}
\phi(r) = A\ \!_2 F_1\left(\frac d 4-\frac 14\sqrt{d^2+4m^2}, \frac d 4 +\frac 14 \sqrt{d^2 +4m^2};\frac d 2; - r^2\right)
\end{equation}
where $A$ indicates the amplitude of the solution.
We now switch to a coordinate $y=1/r$ to study the asymptotic form of the field by expanding near $y\simeq 0$. In this coordinate system we have that  
\begin{equation}
ds ^2 = - \frac {dt^2 }{y^2}- dt^2 + \frac{dy^2}{y^2(1+y^2)}+ \frac{ d \Omega^2 }{y^2}\simeq -\frac {dt^2}{y^2} + \frac {dy^2} {y^2} + \frac{d\Omega^2 }{y^2} 
\end{equation}
So zooming into any small region of the sphere on the boundary $y=0$ we have an asymptotic  form of the metric that matches the usual Poincare slicing of AdS. In such a coordinate system the asymptotic growth or decay  of $\phi(y)$ in the $y$ coordinate is polynomial, proportional to $y^{\Delta_{\pm}}$ and can be matched to the usual dictionary for a flat slicing, where $\Delta_{\pm} = \frac d2 \pm \frac 12\sqrt{d^2 + 4m^2}$. We have made the match 
$
\Delta_+= \Delta,
$
 the operator dimension for irrelevant perturbations. For relevant perturbations we get a choice. 
  
 Reading the coefficients of this expansion has the same interpretation as in flat space: one is a source and the other one is the response. Writing this as
\begin{equation}
\phi(y) \simeq A (f_+ y^{\Delta_+}+f_- y^{\Delta_-})
\end{equation}
we find that $f_+ = \Gamma(d/2) \Gamma(d/2-\Delta_+)/ \Gamma( 1/2 (\Delta_-))^2$, and $f_-$ is the same expression with $\Delta_+$ replaced by $\Delta_-$.
We now use
 \begin{equation}
 \Delta = \Delta_+
 \end{equation}
 in what follows to distinguish between vev and source, although we will find the answer is symmetric in this choice.

 The relation between source and vacuum expectation value is then
\begin{equation}
f_+= \frac{ \Gamma(\frac d2 -\Delta) \Gamma(\frac 12 \Delta)^2}{\Gamma(\Delta-\frac d2) \Gamma(\frac d2 -\frac {\Delta}2)^2} f_-\label{eq:AdSf}
\end{equation}
We have artificially chosen $\Delta=\Delta_+$ over $\Delta _-$ to indicate the vacuum expectation value versus the source as one would do for irrelevant perturbations, but since the expressions for $f_+$ and $f_-$ are symmetric in the exchange of $\Delta_+$ and $\Delta_-$, we can eliminate the distinction in equation \eqref{eq:AdSf}. Notice that this relation seems to be completely independent of the normalization of the field $\phi$. We will explain how to get the correct normalization later.

\subsubsection{The conformal field theory computation}

The basic question for the conformal field theory computation is how does one compute the two point function on the cylinder. Since the cylinder results from a Weyl rescaling of the plane, the two point functions are related to each other in a standard way. The Weyl rescaling is as follows
\begin{equation}
ds^2 =d\vec x^2=  r^2 \left( \frac {dr^2}{r^2}+d \Omega_{d-1}^2\right) \to d\tau^2 + d\Omega_{d-1}^2
\end{equation}
which uses a Weyl factor of $r^2$ (the rescaling of units is by a factor of $r= \exp(\tau)$). As {a} primary field of conformal dimension $\Delta$,  $\CO(x)$ will need to be rescaled by $\CO(\theta, r)\simeq r^\Delta \CO(x)$ to translate to the new rescaled metric. For the  two point functions this means that
\begin{equation}
\langle \CO(\tau_1, \theta_1) \CO(\tau_2,\theta_2) \rangle_{cyl} = \frac{\exp (\Delta \tau_1)\exp( \Delta \tau_2)}{|x_1-x_2|^{2\Delta}}= \frac 1{\left(\exp[ (\tau_1-\tau_2)] + \exp[(\tau_2-\tau_1) ] - 2 \cos(\theta_{rel})\right)^\Delta}
\end{equation}
where $\theta_{rel}$ is the angle computed between the unit vectors $\hat x_1, \hat x_2$ in standard cartesian coordinates. If we choose $\hat x_1$ to be fixed, and at the north pole, the angle $\theta_{rel}$ is the polar angle of the insertion of $\CO$ over which we will integrate. Since the answer only depends on the {the difference of the times,} $\tau_2-\tau_1$, the end result  is time translation invariant. Notice that we have used  throughout conformal primaries that are unit normalized in the Zamolodchikov metric. 

Now we need to integrate over the angles and the relative time $\tau$. Our expression for $C_\Delta$ reduces to the following definite double integral
\begin{eqnarray}
C_\Delta&=&\int_{-\infty} ^\infty d\tau \int_0^\pi d \theta \sin^{d-2}\theta Vol(S_{d-2}) \frac{ 1}{2^\Delta(\cosh \tau - \cos\theta)^\Delta}\\
&=&  2^{1-\Delta} Vol(S_{d-2}) \int_1^\infty du \frac 1{\sqrt{u^2-1}} \int_{-1}^1 dv (1-v^2)^{\frac{d-3}{2}} [u-v]^{-\Delta}\label{eq:int}
\end{eqnarray} 
where we have changed variables to $u =\cosh \tau$ and $v= \cos \theta$. For the integral to converge absolutely, we need that $0<2 \Delta<d$, but once we find an analytic formula for arbitrary $0<2 \Delta<d$ we can analytically continue it for all values of $\Delta,d$. The volume of spheres can be computed in arbitrary dimensions as is done in dimensional regularization, so we also get an analytic answer for the variable $d$ itself.
Any answer we get can therefore be interpreted as one would in a real space dimensional regularization  formalism, where we keep the operator dimension fixed but arbitrary, but where we allow the dimension of space to vary.
 The final answer we get is
\begin{equation}
C_\Delta= \pi^{\frac{(d+1)}2}{ 2^{1-\Delta}} \left[\frac{\Gamma(\frac d2 - \Delta) \Gamma(\frac \Delta 2)}{  \Gamma (\frac d2- \frac\Delta 2)^2\Gamma(\frac 12 +\frac \Delta 2)} \right]\label{eq:CFT2pt}
\end{equation}

\subsubsection{Divergences}

On comparing the answers for the AdS and CFT calculation, equations \eqref{eq:AdSf} and \eqref{eq:CFT2pt} seem to be completely different. But here we need to be careful about normalizations of the operator $\CO$ in the conformal field theory and the corresponding fields in the gravity formulation. We should compare the Green's function of the field $\phi$ in gravity and take it to the boundary to match the two point function one expects in the CFT dual. The correct normalization factor that does so can be found in equation  (A.10) in \cite{Berenstein:1998ij}. Naively, it seems that we just need to multiply the result from equation \eqref{eq:CFT2pt} by $\frac{\Gamma(\Delta)}{2 \pi^{\frac d2} \Gamma (\Delta -\frac d2 +1)} $ and then we might expect 
\begin{equation}
\frac {f_+}{f_-} \simeq \frac{\Gamma(\Delta)}{2 \pi^{\frac d2} \Gamma (\Delta -\frac d2 +1)} C_\Delta.
\end{equation}
However, if we compare the ratio of the left hand side to the right hand side  we get that the ratio of the two is  given by
\begin{equation}
\frac {f_+}{f_-} \left( \frac{\Gamma(\Delta)}{2 \pi^{\frac d2} \Gamma (\Delta -\frac d2 +1)} C_\Delta\right)^{-1}= 2\Delta-d = \Delta_+-\Delta_-
\end{equation}
Happily, this extra factor is exactly what is predicted from the work \cite{Marolf:2004fy} (in particular, eq. 4.24). See also \cite{Freedman:1998tz,Klebanov:1999tb}.
 This is because one needs to add a counter-term to the action of the scalar field when one uses a geometric regulator in order to have a well defined boundary condition in gravity.

We see then that the gravity answer and the field theory answer match each other exactly, for arbitrary $d,\Delta$ once the known normalization issues are dealt with carefully. Now we want to 
interpret  the end result $C_\Delta$ itself.

{The expression we found has singularities at specific values of $\Delta$. These arise from poles in the $\Gamma$ function, which occur when 
$(d/2 - \Delta)$ is a negative integer. However, these poles are cancelled when $(d - \Delta)/2$ is a negative integer, because we then have a double pole in the denominator. For both of these conditions to be
true simultaneously, we need both $d$ and $\Delta$ to be even, and furthermore $\Delta\geq d$.} The origin of such poles is from the UV structure of the integral \eqref{eq:onep}. The singular integral (evaluated at $x=0$) is of the form
\begin{equation}
A_{sing} = \int_0^\epsilon  {d^d y} \, y^{-2\Delta} \propto \int_0^\epsilon dy \, y^{d-1-2\Delta} \simeq \int_{1/\epsilon}^\infty  dp \, p^{d-1} (p^2+m^2)^{-g}\label{eq:dimreg}
\end{equation}
{where $g = d-\Delta$ and in the last step we introduced} a momentum like variable $p=1/y$ and a mass $m$ infrared regulator to render it into a familiar form for dimensional regularization integrals that would arise from Feynman diagrams. Singularities 
on the right hand side arise in dimensional regularization in the UV whenever {there are} logarithmic subdivergences. This {can be seen by}  factorizing $p^2+m^2= p^2(1+ m^2/p^2)$ and expanding in power series in $m^2$. Only when $d-1-2g-2k = -1$ for some  non-negative integer $k$ do we expect {a logarithmic} singularity. In our case, with $-g = \Delta-d$, the condition for such a logarithmic singularity is that  $- g= \Delta-d =-\frac d2+k$, which is exactly the same condition as we found for there to exist poles in the numerator of equation \eqref{eq:CFT2pt}. The first such singularity arises when $\Delta = d/2$. Beyond that, the integral in equation \eqref{eq:int} is not convergent, but is rendered finite in the dimensional regularization spirit. Notice that this was never really an issue in the gravitational computation, since the final answer depended only the asymptotic expansion of hypergeometric functions and we never had to do an integral. The presence of singularities in gravity has to do with the fact that when $\Delta_+-\Delta_-$ is twice an integer, then the two linearly independent solutions to the  hypergeometric equation near $y=0$ have power series expansions where the exponents of $y$ match between the two. Such singularities are resolved by taking a limit which produces an additional logarithm between the two solutions. 
We should take this match to mean that the AdS gravity computation already {\em knows} about dimensional regularization. 

Another interesting value for $\Delta$ is when we take $\Delta\to d$. The denominator will have a double pole that will always render the number $C_{\Delta=d}=0$. This is exactly as expected for a marginal operator in a conformal field theory: it should move us to a near conformal point where all one point functions of non-trivial local operators vanish.

\section{The energy of a quench}

After concluding that the AdS and CFT calculation really {did give} the same answer for a constant perturbation we want to understand the energy stored in such a solution. This needs to be done carefully, because as we have seen divergences can appear. Under such circumstances, we should compare the new state to the vacuum state in the absence of a perturbation and ask if we get a finite answer for the energy. {That} is, we need to take the state and quench the dynamics to the unperturbed theory. In that setup one can compute the energy unambiguously.

We would also like to have  a better understanding of the origin of the divergences in field theory, to understand how one can regulate the UV to create various states we {might} be interested in. For this task we will now do a Hamiltonian analysis. Although in principle one could use a three point function including the stress tensor and integrate, performing a Hamiltonian analysis will both be simpler and more illuminating as to what is the physics of these situations. Also, it is more easily adaptable to a real time situation.

\subsubsection{A Hamiltonian approach}

The perturbation we have discussed in the action takes the Euclidean action $S\to S+\alpha \int \CO$. When thinking in terms of the Hamiltonian on a sphere, we need to take
\begin{equation}
H\to H+ \alpha \int d\Omega' \CO(\theta')
\end{equation}
and we think of it as a new time independent Hamiltonian.
When we think of using $\alpha$ as a perturbation expansion {parameter}, we need to know the action of $\int d\Omega' \CO(\theta')$ on the ground state of the Hamiltonian $\CO(\theta')\ket 0$. This is actually encoded in the two point function we computed. 
Consider the time ordered two point function with $\tau_1>\tau_2$
\begin{eqnarray}
\langle \CO(\tau_1, \theta_1) \CO(\tau_2,\theta_2) \rangle_{cyl}&=&\frac 1{\left(\exp[ (\tau_1-\tau_2)] + \exp[(\tau_2-\tau_1) ] - 2 \cos(\theta_{rel})\right)^\Delta}\\
&=& \sum_s \langle 0| \CO( \theta_1) \exp({- H\tau_1}) \ket s\bra s \exp(H\tau_2) \CO(\theta_2)|0 \rangle \label{eq:Ham2pt}\\
&=& \sum_s \exp(- E_s (\tau_1-\tau_2)) \bra 0 \CO( \theta_2) \ket s \bra s \CO(\theta_1) \ket 0
\end{eqnarray}
where $s$ is a complete basis that diagonalizes the Hamiltonian $H$ and we have written the operators $\CO(\tau)\simeq \exp(H \tau) \CO(0) \exp(-H\tau)$ as corresponds to the Schrodinger picture. The states $\ket s$ that can contribute are those that are related to $\CO$ by the operator-state correspondence: the primary state of $\CO$ and it's descendants. When we integrate over the sphere, only {the} descendants that are spherically invariant can survive. For a primary $\CO(0)$, these are the descendants {given by} $(\partial_\mu\partial^\mu)^k \CO(0)$. The normalized states corresponding to these descendants will have energy (dimension) $\Delta+2k$, and are unique for each $k$. We will label them by $\Delta+2k$. We are interested in computing the amplitudes
\begin{equation}
A_{\Delta+2k}= \bra{\Delta +2 k } \int d\Omega'\CO(\theta')\ket 0\label{eq:ampdef}
\end{equation}
These amplitudes can be read from equation \eqref{eq:Ham2pt} by integration over $\theta_1, \theta_2$. Indeed, we find that 
\begin{eqnarray}
\int {d^{d-1} \Omega} \langle \CO(\tau_1, \theta_1) \CO(\tau_2,\theta_2) \rangle_{cyl}& &=2^{-\Delta} Vol(S_{d-2}) \int_{-1}^1 dv (1-v^2)^{\frac{d-3}{2}} [\cosh(\tau)-v]^{-\Delta}\\
&=&{\pi^{\frac d 2}} 2^{1-\Delta} \cosh[\tau]^{-\Delta} \ _2\tilde F_1[\frac \Delta 2, \frac{1+\Delta}2;\frac d2 ; \cosh^{-2}(\tau)]\label{eq:angint}\\
&=& M \sum |A_{\Delta+2k}|^2 \exp[(-\Delta -2k)\tau]
\end{eqnarray}
where $\tau=\tau_1-\tau_2$ and {$_2\tilde F_1$} is the regularized hypergeometric function.
From this expression further integration over $\Omega_1$ is trivial: it gives the volume of the sphere $Vol(S^{d-1})$. We want to expand this in powers of $\exp(-\tau)$. To do this we use
the expression $\cosh(\tau) = \exp(\tau)( 1+\exp(-2\tau))/2$, and therefore 
\begin{equation}
\cosh^{-a} (\tau) = \exp(- a\tau) 2^{a} [1+\exp(-2\tau)]^{-a} = \sum_{n=0}^\infty 2^{a}\exp(-a \tau -2 n\tau) (-1)^n\frac{\Gamma[a+n]}{n! \Gamma[a]}   
\end{equation}
Inserting this expression into the power series of the hypergeometric function appearing in \eqref{eq:angint} gives us our desired expansion.
Apart from common factors to all the amplitudes $A_{\Delta+2k}$ (which are trivially computed for $k=0$) we are in the end only interested in the $k$ dependence of the amplitude itself. After a bit of algebra one 
finds that
\begin{equation}
|A_{\Delta+2k}|^2 \propto \frac{\Gamma[k+\Delta] \Gamma[\Delta -\frac d2 +k+1]}{\Gamma[1+\Delta-\frac d2]^2 \Gamma[k+\frac d2] k!}\label{eq:kdep}
\end{equation}
and to normalize we have that 
\begin{equation}
|A_{\Delta}|^2= [Vol( S^{d-1})]^2
\end{equation}
For these amplitudes to make sense quantum mechanically, their squares have to be positive numbers. This implies that none of the $\Gamma$ functions in the numerator can be negative. {The condition for that to happen is that the argument of the $\Gamma$ function in the numerator must positive and therefore $\Delta\geq \frac d2 -1$,} which is the usual unitary condition for scalar primary fields. Also, at saturation $\Delta= d/2-1$ we have a free field and then the higher amplitudes vanish $A_{k>0}/A_0=0$. This is reflected in the fact that $\partial_\mu\partial^\mu\phi=0$ is the free field equation of motion.

We are interested in comparing our results to the AdS setup. In the CFT side this usually corresponds to a large $N$ field theory. If the primary fields we are considering are single trace operators, they give rise to an approximate Fock space of states of multitraces, whose anomalous dimension is the sum of the individual traces plus corrections of order $1/N^2$ from non-planar diagrams. In the large $N$ limit we can ignore these corrections, 
 so we want to imagine that the operator insertion of $\CO$ is a linear combination of raising and lowering operators $\int d\Omega \CO(\theta)\simeq \sum A_{\Delta+2k} a^\dagger_{2k+\Delta}+A_{\Delta+2k} a_{2k+\Delta}$ with $[a,a^\dagger]=1$. In such a situation we can write the perturbed Hamiltonian in terms of the free field representation of the Fock space in  the following form
 \begin{equation}
 H+\delta H= \sum E_s a_s^\dagger a_s + \alpha(\sum A_{\Delta+2k} a^\dagger_{2k+\Delta}+A_{\Delta+2k} a_{2k+\Delta}) +O(1/N^2) a^\dagger a^\dagger a a+\dots \label{eq:hamp}
\end{equation}
Indeed, when we work in perturbation theory, if this Fock space exists or not is immaterial, as the expectation value of the energy for a first order perturbation will only depend on the amplitudes we have computed already. It is for states that do not differ infinitesimally from the ground state that we need to be careful about this and this Fock space representation becomes very useful.

When we computed using conformal perturbation theory abstractly, we were considering the vacuum state of the Hamiltonian in equation \eqref{eq:hamp} to first order in $\alpha$. We write this as
\begin{equation}
\ket 0 _\alpha = \ket 0 + \alpha \ket 1
\end{equation}
and we want to compute the value of the energy for the unperturbed Hamiltonian for this new state. This is what quenching the system to the unperturbed theory does for us. We find that
\begin{equation}
\bra {0_\alpha} H\ket 0 _\alpha = \alpha^2 \bra 1 H \ket 1
\end{equation}
Now, we can use the expression \eqref{eq:hamp} to compute the state $\ket 0_\alpha$. Indeed, we find that we can do much better than infinitesimal values of $\alpha$. What we can do is realize that if we ignore the subleading pieces in $N$ then the ground state for $H+\delta H$ is a coherent state for the independent harmonic oscillators $a^\dagger_{2k+\Delta}$. Such a coherent state is of the form
\begin{equation}
\ket 0_\alpha = {\cal N} \exp( \sum \beta_{2k+\Delta} a^\dagger_{2k+\Delta}) \ket 0\label{eq:cohs}
\end{equation}
For such a state we have that 
\begin{equation}
\langle H+\delta H \rangle = \sum (2k+\Delta) |\beta_{2k+\Delta}|^2 + \alpha  \beta_{2k+\Delta} A_{2k+\delta}+\alpha \beta^*_{2k+\Delta}A_{2k+\delta}\label{eq:corrham}
\end{equation}
and the energy is minimized by 
\begin{equation}
\beta_{2k+\Delta} = - \alpha \frac{A_{2k+\Delta}}{2k+\Delta}
\end{equation}
Once we have this information, we can compute the energy of the state  in the unperturbed setup and the expectation value of $\CO$ (which we integrate over the sphere). We find that 
\begin{equation}
\vev H = \sum (2k+ \Delta) |\beta_{2k+\Delta}|^2= \alpha^ 2\sum \frac{|A_{2k+\Delta}|^2}{2k+\Delta}\label{eq:com}
\end{equation}
\begin{equation}
\vev \CO  \simeq \sum 2 A_{k+2\Delta} \beta_{2k+\Delta} \simeq -2 \alpha \sum \frac{|A_{2k+\Delta}|^2 }{2k+\Delta}\label{eq:com2}
\end{equation}
so that in general 
\begin{equation}
\vev H \simeq - \frac{\alpha \vev \CO}{2}
\end{equation}
{That} is, the integrated one point function of the operator $\CO$ over the sphere and the strength of the perturbation is enough to tell us the value of the energy of the state. 
For both of these to be well defined, we need that the sum appearing in \eqref{eq:com} is actually finite. Notice that this matches the Ward identity for gravity \cite{Buchel:2012gw} integrated adiabatically  (for a more general treatment in holographic setups see \cite{Bianchi:2001kw}).

\subsubsection{Amplitude Asymptotics,  divergences and general quenches }

Our purpose now is to understand in more detail the sum appearing in \eqref{eq:com} and \eqref{eq:com2}. We are interested in the convergence and asymptotic values for the terms in the series, {that} is, we want to understand the large $k$ limit. This can be read from equation \eqref{eq:kdep} by using Stirlings approximation $\log \Gamma[t+1]\simeq (t) \log(t) - (t)$ in the large $t$ limit. We find that after using this approximation on all terms that depend on $k$, that 
\begin{eqnarray}
\log( A_{2k+\Delta}^2) &\simeq & (k+\Delta -1) \log(k+\Delta -1) +  (k+\Delta -d/2) \log(k+\Delta -d/2)\\
&&- (k+d/2-1) \log(k+d/2-1)- k\log(k)+O(1)
\\&\simeq&( \Delta-1 +\Delta -\frac d2 -(\frac d 2-1)) \log k = (2\Delta-d)\log k \label{eq:stir}
\end{eqnarray}
So that the sum is bounded by a power law in $k$
\begin{equation}
\sum \frac{|A_{2k+\Delta}|^2}{2k+\Delta}\simeq \sum \frac{1}{k^{ d +1-2\Delta}}
\end{equation}
Again, we see that convergence of the sum requires $2\Delta -d <0$. This is the condition to have a finite vacuum expectation value of both the energy and the operator $\CO$.  If we consider instead the $L^2$ norm of the state, the norm is finite so long as $d+2-2\Delta>1$, {that} is, so long as $\Delta<(d+1)/2$. The divergence in the window $d/2\leq \Delta <(d+1)/2$ is associated {with} the unboundedness of the Hamiltonian, not to the infinite norm of the state. 

In general we can use higher order approximations to find subleading terms in the expression \eqref{eq:stir}. Such approximations will give that $A_{2k+\Delta}$ will have a polynomial expression with leading term  as above, with power corrections in $1/k$. Only a finite number of such corrections lead to divergent sums, so the problem of evaluating $\langle \cal O\rangle $ can be dealt with {using} a finite number of substractions of UV divergences. In this sense, we can renormalize the answer with a finite number of counterterms. A particularly useful regulator to make the sum finite is to choose to modify $A_{2k+\Delta} \to A_{2k+\Delta} \exp (- \epsilon (2k+\Delta))$. 
This is like inserting the operator $\CO$ at time $t=0$ in the Euclidean cylinder and evolving it in Euclidean time for a time $\epsilon$. Because the growth of the coefficients is polynomial in $k$, any such exponential will
render the sum finite. We can trade the divergences in the sums for powers of $1/\epsilon$ and then take the limit $\epsilon\to 0$ of the regulated answer. This is beyond the scope of the present paper.

Notice that we can also analyze more general quenches from studying equation \eqref{eq:corrham}. All we have to do is make $\alpha$ time dependent. The general problem can then be analyzed in terms of linearly driven harmonic oscillators, one for each $a^\dagger, a$ pair. Since the driving force is linear in raising and lowering operators, the final state will always be a coherent state as in equation \eqref{eq:cohs} for some $\beta$ which is the linear response to the 
source. The differential equation,  derived from the Schrodinger equation applied to a time dependent coherent state, is the following
 \begin{equation}
i \dot \beta_{2k+\Delta} (t)= (2k+\Delta) \beta_{2k+\Delta} + \alpha(t) A_{2k+\Delta}\label{eq:deq}
\end{equation}
The solution is given by
\begin{equation}
\beta_{2k+\Delta}(t) =  \beta_{2k+\Delta}(0)\exp (-i \omega t) + A_{2k+\Delta}\int^\infty_0 dt' \alpha(t') \theta(t-t') \exp(-i \omega(t-t')) 
\end{equation}
with $\omega= 2k+\Delta$ the frequency of the oscillator.

Consider the case that $\alpha$ only acts over a finite amount of time between $0, \tau$ and that we start in the vacuum. After the time $\tau$ the motion for $\beta$ will be trivial, and the amplitude will be given by
\begin{equation}
\beta_{2k+\Delta}(\tau) = A_{2k+\Delta}\exp(-i (2k+\Delta ) \tau) \int^\tau_0 dt' \alpha(t')  \exp(i \omega t')
\end{equation} 
and all of these numbers can be obtained from the Fourier transform of $\alpha(t)$. Notice that these responses are always correct in the infinitesimal $\alpha$ regime, as can be derived using time dependent perturbation theory. What is interesting is that in the large $N$ limit they are also valid for $\alpha(t)$ that is not infinitesimal, so long as the $O(1/N)$ corrections can still be neglected. One can also compute the energy of such processes. In particular, so  long as $\Delta <d/2$, any such experiment with  bounded $\alpha(t)$ will give a finite answer.

The simplest such experiment is to take $\alpha$ constant during a small interval $\tau=\delta t<<1$. For modes with small $\omega$, {that} is, those such that $\omega \delta t <1$, we then have that
\begin{equation}
\beta_{2k+\Delta}(\tau)\simeq A_{2k+\Delta} \alpha \delta t 
\end{equation}
While for those modes such that $\omega \delta t>1$, we get that 
\begin{equation}
|\beta_{2k+\Delta} (\tau)| \simeq  \alpha\frac{A_{2k+\Delta}}{\omega}
\end{equation} 
When we compute the energy of such a configuration, we need to divide the sum between high frequency and low frequency modes. The energy goes as 
\begin{equation}
E\simeq \sum \omega |\beta_{2k+\Delta}|^2 \simeq \int_0^{1/(2\delta t) } dk \omega |A_{2k+\Delta} \alpha \delta t|^2 +\int_{1/(2 \delta t)}^{\infty } dk  \frac{|\alpha A_{2k+\Delta}|^2}{\omega}
\end{equation}
now we use the fact that $|A_{2k+\Delta}|^2 \simeq k^{2\Delta -d}$ and that $\omega\propto k$ to find that 
\begin{equation}
E \simeq |\alpha|^2 (\delta t)^{d-2\Delta}
\end{equation}
which shows an interesting power law for the energy deposited into the system. 
One can similarly argue that the one point function of $\CO(\tau)$ scales as $\alpha (\delta t)^{d-2\Delta}$: for the slow modes, the sum is proportional to $\sum A_{2k+\Delta}^2\alpha\delta t$, while for the fast modes one can argue that they have random phases and don't contribute to $\CO(\tau)$.

If we want to study the case $\Delta \geq d/2$, divergences arise, so we need to choose an $\alpha(t)$ that is smooth enough that the high energy modes are not excited in the process because they are adiabatic, but if we scale that into a $\delta t$ window, the adiabatic modes are going to be those such $\omega \delta t > 10$, {let's} say. 
Then for these modes we take $\beta\simeq 0$, and then the estimate is also as above. For $\Delta= d/2$, in an abrupt quench one obtains a logarithmic singularity rather than power law, coming from the UV modes. 
This matches the results in \cite{Das:2014jna} and gives a reason for their universality as arising from the universality of 2-point functions in  conformal perturbation theory. Essentially, the nature of the singularities that arise is that 
the amplitudes to generate descendants are larger than amplitudes to generate primaries, so the details of the cutoff matter.

Here is another simple way to understand the scaling for the one point function of the operator $\CO(\tau)$. The idea is that we need to do an integral similar to $\int d^d x \CO(\tau)\alpha(x) \CO(x)$, but which takes into account causality of the perturbation relative to the response. If we only turn on the perturbation by a small amount of time $\delta t$, the backwards lightcone volume to the insertion of an operator at $\tau=\delta t $ is of order $\delta t^d$, and this finite volume serves as an infrared regulator,  while the two point function that is being integrated is of order $\delta t ^{-2 \Delta}$. When we combine these two pieces of information we get a result proportional to $\delta t ^{d-2\Delta}$, which again is finite for $\Delta < d/2$ and otherwise has a singularity in the corresponding integral. Similarly, the energy density would be an integral of the three point function $T \CO \CO \simeq \delta t^{-2 \Delta-  d} $ times the volume of the past lightcone squared which is again proportional to $\delta t ^{2d}$, giving an answer with the scaling we have already found. The additional corrections would involve an extra insertion of $\CO$ and the volume of the past lightcone, so they scale as $\delta t^{d-\Delta}$, multiplied by the amplitude of the perturbation. This lets us recover the scalings of the energy \cite{Das:2014jna} in full generality.

\subsubsection{A note on renormalization}

So far we have described our experiment as doing a time dependent profile for $\alpha(t)$ such that $\alpha(t)=0$ for $t>\tau$. Under such an experiment, we can control the outcome of the operations we have described and we obtain the scaling relations that we want. If on the other hand we want to measure the operator $\CO(t,\theta)$ for some $t<\tau$, we need to be more careful. This is where we need a better prescription for subtracting divergences: as we have seen, in the presence of a constant value for  $\alpha$ we already have divergences for $\Delta\geq d/2$. 
Under the usual rules of effective field theory, all UV divergences correcting $\langle \CO(t,\theta) \rangle $ should be polynomial in the (perturbed) coupling constants and their derivatives. 
 Moreover, if a covariant regulator exists, it suggests that the number of such derivatives should be even.
Since we are working to linear order in $\alpha(t)$, these can only depend 
linearly on $\alpha(t)$ and it's time derivatives. Another object that can show up regularly is the curvature of the background metric in which we are doing conformal field theory. That is, we can have expressions of the form $\partial_t^k \alpha(t) R^s$ and $(\partial_t^k \alpha \partial_t^\ell \alpha ) R^s$ appearing as counterterms in the effective action. These are needed if we want to compute the energy during the quench.  Although in principle we can also get covariant derivatives acting on the curvature tensor $R$, these vanish on the cylinder.
The counterterms are particularly important in the case of logarithmic divergences, as these control the renormalization group. Furthermore, the logarithmic divergences are usually the only divergences that are immediately visible in dimensional regularization.
It is also the case that there are logarithmic enhancements of the maximum value of $\CO(t,\theta)$ during the quench \cite{Das:2014jna} and these will be captured by such logarithmic divergences. 

We need to identify when such logarithmic divergences can be present. In particular, we want to do a subtraction of the adiabatic modes (which do contribute divergences) to the one point function of $\CO(\theta, t)$ at times $t<\tau$. To undertake such a procedure, we want to solve equation \eqref{eq:deq} recursively for the adiabatic modes (those high $k$). We do this by taking 
\begin{equation}
\beta_{2k+\Delta}(t) = -\alpha(t)\frac{A_{2k+\Delta}}{2k+\Delta} + \beta_{2k+\Delta}^1(t) +\beta_{2k+\Delta}^2(t) +\dots 
\end{equation}
where we determine the $\beta_i(t)$ recursively for high $k$ by substituting $\beta_{2k+\Delta}(t)$ as above in the differential equation. The solution we have written is correct to zeroth order, and we then write the next term as follows
\begin{equation}
-i\dot \alpha(t)\frac{A_{2k+\Delta}}{2k+\Delta} = (2k+\Delta) \beta_{2k+\Delta}^1(t)
\end{equation}
and in general 
\begin{equation}
i \dot \beta_{2k+\Delta}^{n-1}(t)= (2k+\Delta) \beta_{2k+\Delta}^n(t) \label{eq:rec}
\end{equation}
This will generate a series in $\frac1{(2k+\Delta)^n}\partial_t^n \alpha(t)$, which is also proportional to $A_{2k+\Delta}$. We then substitute this solution into the expectation value of $\CO(t, \theta)$, where we get an expression of the form
\begin{equation}
\vev{\CO(t)} \simeq \sum_{k, \ell} |A_{2k+\delta}|^2 \frac{c_\ell}{(2k+\Delta)^{\ell+1}}\partial_t^\ell \alpha(t) \simeq \int dk\sum_{\ell}\frac{1}{k^{2\Delta-d}} \frac{c_\ell}{(2k+\Delta)^{\ell+1}}\partial_t^\ell \alpha(t)
\end{equation}
The right hand side has a logarithmic divergence when $2\Delta-d + \ell=0$. Notice that this divergence arises from the combination $\beta+\beta^*$, so the terms with odd derivatives vanish because of the factors of $i$
in equation \eqref{eq:rec}. Thus, such logarithmic divergences will only be present when $\ell $ is even. This matches the behavior we expect when we have a covariant regulator.
This translates to $\Delta= d/2 +k$, where $k$ is an integer. Notice that this is the same condition that we need to obtain a pole in the numerator of the Gamma function in equation \eqref{eq:int}. We see that such logarithmic divergences are exactly captured by dimensional regularization. As a logarithmic divergence, it needs to be of the form $\log( \Lambda_{UV}/\Lambda_{IR})= \log(\Lambda_{UV}/\mu)+\log(\mu/\Lambda_{IR})$. In our case, the $IR$ limit is formally set by the radius of the sphere, while the UV is determined by how we choose to work precisely with the cutoff. The counterterm is the infinite term $ \log(\Lambda_{UV}/\mu)$, but the finite term depends on the intermediate scale $\mu$, which is also usually taken to be a UV scale which is finite. This lets us consider the Lorentzian limit by taking a small region of the sphere and to work with $\delta t$ as our infrared cutoff: only the adiabatic modes should be treated in the way we described above. Then the logarithmic term scales as $\log((\mu\delta t))\partial_t^{2\Delta-d} \alpha(t) $. These logarithmic terms are exactly as written in \cite{Das:2014jna}. Notice that after the quench, we have that $\alpha(t)=0$ and all of it's derivatives are zero, so no counterterms are needed at that time. We only need the pulse $\alpha(t)$ to be smooth enough so that the state we produce has finite energy.

\section{Conclusion}

In this paper we have shown how to do conformal perturbation theory on the cylinder rather than in flat space. The main reason to do so was to use a physical infrared regulator in order to 
understand the process of renormalization of UV divergences in a more controlled setting. We showed moreover that the results that are found using AdS calculations actually match a notion of dimensional regularization where the dimension of the  perturbation operator stays fixed. In this sense the AdS geometry knows about dimensional regularization as a regulator. This is an interesting observation that merits closer attention.
In particular, it suggests that one can try a real space dimensional regularization approach to study perturbations of conformal field theory. 

We then showed that one could treat in detail also a time dependent quench, and not only where we able to find the energy after a quench, but we also were able to understand scalings that have been observed before for fast quenches.  Our calculations show in what sense they are universal. They only depend on the two point function of the perturbation. The singularities that arise can be understood in detail in the Hamiltonian formulation we have pursued, and they arise from amplitudes to excite descendants increasing with energy, or just not decaying fast enough. In this way they are sensitive to the UV cutoff associated to a pulse quench: the Fourier transform of the pulse shape needs to decay sufficiently fast at infinity to compensate for the increasing amplitudes to produce descendants. We were also able to explain some logarithmic enhancements for the vacuum expectation values of operators  during the process of the quench that can be understood in terms of renormalizing the theory to first order in the perturbation. Understanding how to do this to higher orders in the perturbation is interesting and should depend on the OPE coefficients of a specific theory.

\acknowledgments

D.B. Would like to thank D. Marolf, R. Myers, J. Polchinski and M. Srednicki for discussions. D. B.  is supported in part by the DOE under grant DE-FG02-91ER40618.

\end{document}